\documentclass{ws-p8-50x6-00}

\begin{document}

\def\sss{\scriptscriptstyle}
\def\barp{{\raise.35ex\hbox{${\sss (}$}}---{\raise.35ex\hbox{${\sss )}$}}}
\def\bdbarp{\hbox{$B_d$\kern-1.4em\raise1.4ex\hbox{\barp}}}
\def\bsbarp{\hbox{$B_s$\kern-1.4em\raise1.4ex\hbox{\barp}}}
\def\dbarp{\hbox{$D$\kern-1.1em\raise1.4ex\hbox{\barp}}}
\def\dcp{D^0_{\sss CP}}
\def\dbar{{\overline{D^0}}}
\def\ks{K_{\sss S}}
\newcommand{\xd}{x_d}
\newcommand{\xs}{x_s}
\newcommand{\bd}{B_d^0}
\newcommand{\bdbar}{\overline{B_d^0}}
\newcommand{\bs}{B_s^0}
\newcommand{\bsbar}{\overline{B_s^0}}
\newcommand{\bu}{B_u^\pm}
\newcommand{\beq}{\begin{equation}}
\newcommand{\eeq}{\end{equation}}
\newcommand{\absvcb}{\vert V_{cb}\vert}
\newcommand{\absvub}{\vert V_{ub}\vert}
\newcommand{\absvtd}{\vert V_{td}\vert}
\newcommand{\absvts}{\vert V_{ts}\vert}
\newcommand{\abseps}{\vert\epsilon\vert}
\newcommand{\epsp}{\epsilon^\prime/\epsilon}
\newcommand{\fbb}{f^2_{B_d}\hat{B}_{B_d}}
\newcommand{\fbbs}{f^2_{B_s}\hat{B}_{B_s}}
\newcommand{\fbd}{f_{B_d}}
\newcommand{\fbs}{f_{B_s}}
\newcommand{\fds}{f_{D_s}}
\def\rly#1{\mathrel{\raise.3ex\hbox{$#1$\kern-.75em\lower1ex\hbox{$\sim$}}}}
\def\lsim{\rly<}
\def\thetaNP{\theta_{\sss NP}}
\def\twida{\tilde{\alpha}}
\def\twidb{\tilde{\beta}}
\def\twidg{\tilde{\gamma}}
\def\BK{B_{\sss K}}
\def\BBd{B_{\sss B_d}}
\def\nn{\nonumber}

\def \epjc#1#2#3{{\it Eur.\ Phys.\ J.}\ C {\bf #1}, #3 (19#2)}
\def \zpc#1#2#3{{\rm Z.~Phys.} C {\bf #1}, #3 (19#2)}
\def \plb#1#2#3{{\rm Phys.~Lett.} B {\bf #1}, #3 (19#2)}
\def \ibj#1#2#3{~#1, #3 (19#2)}
\def \prl#1#2#3{{\rm Phys.~Rev.~Lett.} {\bf #1}, #3 (19#2)}
\def \prd#1#2#3{{\rm Phys.~Rev.} D {\bf #1}, #3 (19#2)} 
\def \newprd#1#2#3{{\it Phys.\ Rev.} D {\bf #1}: #3 (19#2)}
\def \npb#1#2#3{{\rm Nucl.~Phys.} B {\bf #1}, #3 (19#2)} 
\def\ijmp#1#2#3{{\rm Int.\ J.\ Mod.\ Phys.} A {\bf #1}, #3 (19#2)}
\def \stone{{\it B Decays}, edited by S. Stone (World Scientific, Singapore,
1994)}

\newread\epsffilein 
\newif\ifepsffileok 
\newif\ifepsfbbfound 
\newif\ifepsfverbose 
\newdimen\epsfxsize 
\newdimen\epsfysize 
\newdimen\epsftsize 
\newdimen\epsfrsize 
\newdimen\epsftmp 
\newdimen\pspoints 
\pspoints=1bp 
\epsfxsize=0pt 
\epsfysize=0pt 
\def\epsfbox#1{\global\def\epsfllx{72}\global\def\epsflly{72}%
 \global\def\epsfurx{540}\global\def\epsfury{720}%
 \def\lbracket{[}\def\testit{#1}\ifx\testit\lbracket
 \let\next=\epsfgetlitbb\else\let\next=\epsfnormal\fi\next{#1}}%
\def\epsfgetlitbb#1#2 #3 #4 #5]#6{\epsfgrab #2 #3 #4 #5 .\\%
 \epsfsetgraph{#6}}%
\def\epsfnormal#1{\epsfgetbb{#1}\epsfsetgraph{#1}}%
\def\epsfgetbb#1{%
%
%
\openin\epsffilein=#1
\ifeof\epsffilein\errmessage{I couldn't open #1, will ignore it}\else
%
%
 {\epsffileoktrue \chardef\other=12
 \def\do##1{\catcode`##1=\other}\dospecials \catcode`\ =10
 \loop
 \read\epsffilein to \epsffileline
 \ifeof\epsffilein\epsffileokfalse\else
%
%
 \expandafter\epsfaux\epsffileline:. \\%
 \fi
 \ifepsffileok\repeat
 \ifepsfbbfound\else
 \ifepsfverbose\message{No bounding box comment in #1; using defaults}\fi\fi
 }\closein\epsffilein\fi}%
%
%
\def\epsfclipstring{}
\def\epsfclipon{\def\epsfclipstring{ clip}}%
\def\epsfclipoff{\def\epsfclipstring{}}%
\def\epsfsetgraph#1{%
 \epsfrsize=\epsfury\pspoints
 \advance\epsfrsize by-\epsflly\pspoints
 \epsftsize=\epsfurx\pspoints
 \advance\epsftsize by-\epsfllx\pspoints
%
%
 \epsfxsize\epsfsize\epsftsize\epsfrsize
 \ifnum\epsfxsize=0 \ifnum\epsfysize=0
 \epsfxsize=\epsftsize \epsfysize=\epsfrsize
 \epsfrsize=0pt
%
%
 \else\epsftmp=\epsftsize \divide\epsftmp\epsfrsize
 \epsfxsize=\epsfysize \multiply\epsfxsize\epsftmp
 \multiply\epsftmp\epsfrsize \advance\epsftsize-\epsftmp
 \epsftmp=\epsfysize
 \loop \advance\epsftsize\epsftsize \divide\epsftmp 2
 \ifnum\epsftmp>0
 \ifnum\epsftsize<\epsfrsize\else
 \advance\epsftsize-\epsfrsize \advance\epsfxsize\epsftmp \fi
 \repeat
 \epsfrsize=0pt
 \fi
 \else \ifnum\epsfysize=0
 \epsftmp=\epsfrsize \divide\epsftmp\epsftsize
 \epsfysize=\epsfxsize \multiply\epsfysize\epsftmp
 \multiply\epsftmp\epsftsize \advance\epsfrsize-\epsftmp
 \epsftmp=\epsfxsize
 \loop \advance\epsfrsize\epsfrsize \divide\epsftmp 2
 \ifnum\epsftmp>0
 \ifnum\epsfrsize<\epsftsize\else
 \advance\epsfrsize-\epsftsize \advance\epsfysize\epsftmp \fi
 \repeat
 \epsfrsize=0pt
 \else
 \epsfrsize=\epsfysize
 \fi
 \fi
%
%
 \ifepsfverbose\message{#1: width=\the\epsfxsize, height=\the\epsfysize}\fi
 \epsftmp=10\epsfxsize \divide\epsftmp\pspoints
 \vbox to\epsfysize{\vfil\hbox to\epsfxsize{%
 \ifnum\epsfrsize=0\relax
 \includegraphics{#1}%
 \else
 \epsfrsize=10\epsfysize \divide\epsfrsize\pspoints
 \includegraphics{#1}%
 \fi
 \hfil}}%
\global\epsfxsize=0pt\global\epsfysize=0pt}%
%
%
 {\catcode`\%=12 \global\let\epsfpercent=
%
%
\long\def\epsfaux#1#2:#3\\{\ifx#1\epsfpercent
 \def\testit{#2}\ifx\testit\epsfbblit
 \epsfgrab #3 . . . \\%
 \epsffileokfalse
 \global\epsfbbfoundtrue
 \fi\else\ifx#1\par\else\epsffileokfalse\fi\fi}%
%
%
\def\epsfempty{}%
\def\epsfgrab #1 #2 #3 #4 #5\\{%
\global\def\epsfllx{#1}\ifx\epsfllx\epsfempty
 \epsfgrab #2 #3 #4 #5 .\\\else
 \global\def\epsflly{#2}%
 \global\def\epsfurx{#3}\global\def\epsfury{#4}\fi}%
%
%
\def\epsfsize#1#2{\epsfxsize}
%
%
\let\epsffile=\epsfbox

\begin{flushright}
UdeM-GPP-TH-00-69 \\
March, 2000 \\
\end{flushright}

\vspace{0.5cm}
\centerline{\bf DISCRETE AMBIGUITIES IN $B$-DECAY CP ASYMMETRIES}
\centerline{\bf AND THE SEARCH FOR NEW PHYSICS\footnote{Invited talk,
    to be published in the Proceedings of PASCOS99, 7th International
    Symposium on Particles, Strings and Cosmology, Lake Tahoe,
    California, Dec. 10-16, 1999} } 
\vspace{1.0cm}

\begin{center}
David London \\
Laboratoire Ren\'e J.-A. L\'evesque, Universit\'e de Montr\'eal,  \\ 
C.P. 6128, succ. centre-ville, Montr\'eal, QC, Canada H3C 3J7 \\
E-mail: london@lps.umontreal.ca
\end{center}

\vspace{0.3cm}

\abstracts{The first measurements of CP violation in the $B$ system
will probably extract $\sin 2\alpha$, $\sin 2\beta$ and $\cos
2\gamma$. Assuming that the CP angles $\alpha$, $\beta$ and $\gamma$
are the interior angles of the unitarity triangle, this determines the
angle set $(\alpha,\beta,\gamma)$ up to a twofold discrete ambiguity.
The presence of this discrete ambiguity can make the discovery of new
physics difficult: if only one of the two solutions is consistent with
constraints from other measurements in the $B$ and $K$ systems, one is
not sure whether new physics is present or not. I present examples of
this situation, and discuss ways to resolve the discrete ambiguity.}

Within the standard model (SM), CP violation is due to the presence of
a nonzero phase in the Cabibbo-Kobayashi-Maskawa (CKM) mixing matrix.
This phase information can be elegantly displayed using the unitarity
triangle, in which the interior (CP-violating) angles are labelled
$\alpha$, $\beta$ and $\gamma$. Constraints on the unitarity triangle
come from several sources ($\absvcb$, $|V_{cb}/V_{ub}|$, $\abseps$,
$\bd$-$\bdbar$ and $\bs$-$\bsbar$ mixing), and the allowed region is
shown in Fig.~\ref{rhoeta1}~\cite{AliLon}. (In the following, I will
refer to this as the ``allowed UT region.'') Note that this region is
still relatively large -- at present the position of the apex of the
triangle is not well-established. This is due principally to the large
theoretical hadronic uncertainties present in some of the measured
quantities.

\begin{figure}
\vskip -1.0truein
\centerline{\epsfxsize 3.5 truein \epsfbox {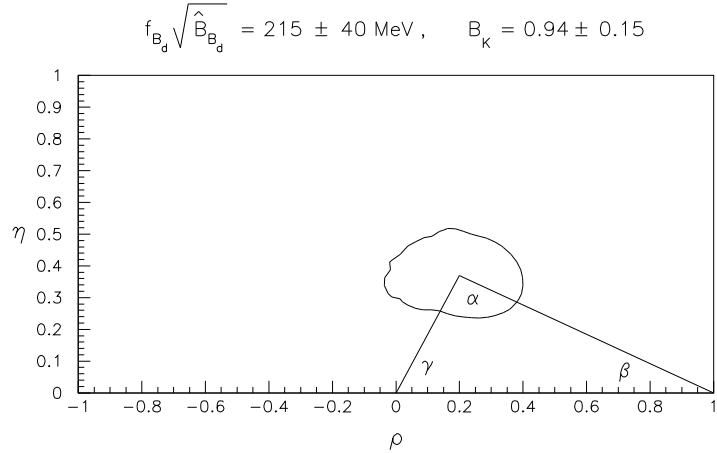}}
\vskip -1.5truein
\caption{Allowed region of the unitarity triangle at 95\% C.L.}
\label{rhoeta1}
\end{figure}

In the coming years, CP-violating rate asymmetries in $B$ decays will
be measured at a variety of machines. These asymmetries probe the
angles $\alpha$, $\beta$ and $\gamma$ {\it without hadronic
  uncertainties}, and will therefore allow us to cleanly reconstruct
the unitarity triangle. If Nature is kind, these measurements may
reveal the presence of new physics. This can happen as follows.

The principal way in which new physics can affect CP asymmetries is by
changing the phase of the neutral $B$-${\bar B}$ mixing
amplitudes~\cite{DLN}, thereby shifting the CP angles from their SM
values. Note, however, that the angles $\alpha$, $\beta$ and $\gamma$
will probably be first probed via CP violation in $\bd(t) \to \pi\pi$
(or $\rho\pi$~\cite{Dalitz}), $\bd(t) \to \Psi\ks$ and $B^\pm \to D
K^\pm$~\cite{BtoDK}, respectively. If there is new physics in
$B$-${\bar B}$ mixing, only the measurements of $\alpha$ and $\beta$
will be affected. Furthermore, they will be affected in opposite
directions~\cite{nirsilv}. That is, instead of extracting $\alpha$ and
$\beta$, the decays will measure $\alpha + \thetaNP$ and $\beta -
\thetaNP$. The upshot is that new physics will {\it not} be discovered
through a violation of the triangle condition $\alpha + \beta + \gamma
= \pi$.

However, new physics {\it can} be found if the triangle constructed
from measurements of the angles is inconsistent with that constructed
from measurements of the sides. Unfortunately, there are some
potential problems with this procedure. First, because the allowed
unitarity-triangle region is still relatively large (see
Fig.~\ref{rhoeta1}), it is conceivable that new physics might be
present, but the triangle constructed from the CP angles $\alpha$,
$\beta$ and $\gamma$ might still lie within the allowed region. Of
course, this would suggest that the new-physics angle $\thetaNP$ is
small. Still, we would like to be able to detect such a new-physics
effect. 

More importantly, even if $\thetaNP$ is large, the
$\alpha$-$\beta$-$\gamma$ triangle may still be within the allowed
region. This is due to the presence of discrete
ambiguities~\cite{disamb}. The point is that we don't actually measure
the CP phases $\alpha$, $\beta$ and $\gamma$. Instead, it is the
functions $\sin 2\alpha$, $\sin 2\beta$ and $\sin^2 \gamma$ (or,
equivalently, $\cos 2\gamma$) which are extracted. And if new physics
is present, the angles probed are in fact $\twida$, $\twidb$ and
$\twidg$, where
\beq
\twida = \alpha + \thetaNP~,~~~~
\twidb = \beta - \thetaNP~,~~~~
\twidg = \gamma~.
\label{twidangles}
\eeq

{}From $\sin 2\twida$, the CP phase $\twida$ can be extracted with a
fourfold ambiguity, and similarly for $\sin 2\twidb$ and $\cos
2\twidg$. Thus, there is a 64-fold ambiguity in the angle set
$(\twida,\twidb,\twidg)$. Of course, in practice, we will assume that
$\twida$, $\twidb$ and $\twidg$ are the interior angles of the
unitarity triangle, i.e.\ (i) they are all of same sign, and (ii)
$|\twida+\twidb+\twidg| = 180^\circ$. (Note: negative CP angles
correspond to a downward-pointing unitarity triangle, which implies
that the bag parameters $\BK$ and/or $\BBd$ are negative. This
scenario is disfavoured theoretically~\cite{signBK}, but should be
checked experimentally.) The key point here is that not all angle sets
satisfy these two conditions. In fact, the measurements of $\sin
2\twida$, $\sin 2\twidb$ and $\cos 2\twidg$ determine
$(\twida,\twidb,\twidg)$ up to a twofold ambiguity~\cite{disamb}.

Given this twofold ambiguity, there are then three possibilities:
\begin{enumerate}

\item Both candidate solutions are consistent with the allowed UT
region.

\item Only one candidate solution is consistent with the allowed UT
region.

\item Neither candidate solution is consistent with the allowed UT
region.

\end{enumerate}
In practice, situation (1) cannot arise. According to the constraints
shown in Fig.~\ref{rhoeta1}, the allowed range for $\beta$ is
$16^\circ \le \beta \le 35^\circ$. The twofold discrete ambiguity has
$\beta \to {\pi\over 2} - \beta$ or $\beta \to -{\pi\over 2} - \beta$.
It is therefore impossible for both solutions to satisfy the
constraint on $\beta$. In addition, situation (3) poses no problem. In
this case we will {\it know} that new physics is present (though we
would still like to know which is the correct solution).

It is situation (2) which is problematic. Is new physics present or
not? Below I give some examples of this scenario.

Suppose that the SM values of the CP angles are $(\alpha,\beta,\gamma)
= (70^\circ,20^\circ,90^\circ)$. This corresponds to a point near the
left-hand side of the allowed UT region.
\begin{itemize}

\item Suppose $\thetaNP = -50^\circ$. The two solutions are then
\begin{eqnarray}
(\twida_1,\twidb_1,\twidg_1) & = & (20^\circ,70^\circ,90^\circ) ~, \nn\\
(\twida_2,\twidb_2,\twidg_2) & = & (70^\circ,20^\circ,90^\circ) ~.
\end{eqnarray}
The first solution is completely inconsistent with the allowed UT
region, but the second solution is consistent. In fact, the second
solution is {\it identical} to the SM solution, even though there is a
large $\thetaNP$!

\item Suppose $\thetaNP = +130^\circ$. In this case the values of the
new-physics-modified CP angles [Eq.~(\ref{twidangles})] are
$(\twida,\twidb,\twidg) = (-160^\circ,-110^\circ,90^\circ)$, which is
not even a triangle. (Since two CP angles have changed sign here, I
refer to this situation as having ``two flips.'') Even in this
situation, however, there are two solutions which form a
triangle. They are
\begin{eqnarray}
(\twida_1,\twidb_1,\twidg_1) & = & (20^\circ,70^\circ,90^\circ) ~, \nn\\
(\twida_2,\twidb_2,\twidg_2) & = & (70^\circ,20^\circ,90^\circ) ~.
\end{eqnarray}
As above, the first solution is inconsistent with the allowed UT
region, but the second solution is consistent.

\item Suppose $\thetaNP = +90^\circ$. Here the values of the
new-physics-modified CP angles are $(\twida,\twidb,\twidg) =
(160^\circ,-70^\circ,90^\circ)$, which is also not a triangle. (Since
one CP angle has changed sign, this situation is referred to as having
``one flip.'') The two solutions which do form a triangle are
\begin{eqnarray}
(\twida_1,\twidb_1,\twidg_1) & = & (-20^\circ,-70^\circ,-90^\circ) ~, \nn\\
(\twida_2,\twidb_2,\twidg_2) & = & (-70^\circ,-20^\circ,-90^\circ) ~.
\end{eqnarray}
As usual, the first solution is inconsistent with the allowed UT
region. The second solution is consistent only if we allow for the
possibility that the unitarity triangle might point down, i.e.\ that
the bag parameter $\BK$ may be negative.

\end{itemize}
{}From the above examples, it is clear that we can categorize the
various solutions according to the number of flips. This is summarized
in Table \ref{fliptable}.

\begin{table}
\hfil
\vbox{\offinterlineskip
\halign{&\vrule#&
\strut\quad#\hfil\quad\cr
\noalign{\hrule}
height2pt&\omit&&\omit&\cr
& Angle(s) flipped && $(\twida_1,\twidb_1,\twidg_1$) & \cr
height2pt&\omit&&\omit&\cr
\noalign{\hrule}
height2pt&\omit&&\omit&\cr
& none && $\left(\twida, \twidb, \twidg \right)$ & \cr
& $\alpha$ && $\left(\twida, \twidb - \pi, \twidg - \pi \right)$ & \cr
& $\beta$ && $\left(\twida - \pi, \twidb, \twidg - \pi\right)$ & \cr
& $\alpha$, $\beta$ && $\left(\twida - \pi, \twidb + \pi, \twidg
\right)$ & \cr
height2pt&\omit&&\omit&\cr
\noalign{\hrule}}}
\caption{Construction of the triangle angle set
$(\twida_1,\twidb_1,\twidg_1)$ in the presence of new physics.}
\label{fliptable}
\end{table}

The above examples demonstrate that it is possible for one of the two
discretely ambiguous solutions to be consistent with the allowed UT
region, even in the presence of a large new-physics phase. In order to
establish whether such large new physics is present or not, it will be
necessary to remove this discrete ambiguity. This can be done if a
different function of $\twida$, $\twidb$ or $\twidg$ is measured. For
example, $\cos 2\twida$ can be obtained through a study of the
time-dependent Dalitz plot for $\bd(t)\to\rho\pi$
decays~\cite{Dalitz}. Similarly, Dalitz-plot analyses of the decays
$\bd(t) \to D^+ D^- \ks$ and $\bd(t) \to D^\pm \pi^\mp \ks$ allow one
to extract the functions $\cos 2\twidb$ and $2(2\twidb + \twidg)$,
respectively~\cite{Charlesetal}. $\cos 2\twidb$ can also be obtained
through a study of $\bd \to \Psi + K \to \Psi + (\pi^- \ell^+ \nu)$,
known as ``cascade mixing''~\cite{cascade}. Finally, $\sin 2\twidg$
can be obtained from $\bs(t)\to D_s^\pm K^\mp$ if the width difference
between the two $B_s$ mass eigenstates is measurable~\cite{IsiBs}.

All of these measurements are quite difficult, but one may be
necessary in order to discover new physics. In all cases, the
knowledge of the additional function of $\twida$, $\twidb$ or $\twidg$
{\it eliminates} the $(\twida_2,\twidb_2,\twidg_2)$ solution. (Recall
that in the examples above, the $(\twida_2,\twidb_2,\twidg_2)$
solution was the one which was consistent with the allowed UT region.)

Does this always work? No: if $\thetaNP$ is close to 0 or $\pi$, then
discrete ambiguity resolution (DAR) will not reveal the presence of
new physics. This can be seen from Table \ref{fliptable}. If $\thetaNP
\approx 0$, then there will be no flips, and DAR will choose the
solution $(\twida,\twidb,\twidg)$, which is close to the SM solution
$(\alpha,\beta,\gamma)$. And if $\thetaNP \approx \pi$, then there
will be two flips, and DAR will choose
$(\twida-\pi,\twidb+\pi,\twidg)$, which again is close to
$(\alpha,\beta,\gamma)$. In both cases, DAR will choose the solution
which is consistent with the allowed UT region, even though (small)
new physics is present.

Note, however, that $\thetaNP \approx 0$ need not be {\it that} small.
For example, suppose that the SM values of the CP angles are
$(\alpha,\beta,\gamma) = (113^\circ,17^\circ,50^\circ)$, and that
$\thetaNP = -20^\circ$. Then the two discretely ambiguous solutions are
\begin{eqnarray}
(\twida_1,\twidb_1,\twidg_1) & = & (93^\circ,37^\circ,50^\circ) ~, \nn\\
(\twida_2,\twidb_2,\twidg_2) & = & (-3^\circ,-127^\circ,-50^\circ) ~.
\end{eqnarray}
Here it is the second solution which is inconsistent with the allowed
UT region. The first solution, which will be chosen by DAR, is still
consistent. Of course, one can reduce the likelihood of this
particular scenario occurring by reducing the allowed UT region.

Finally, in most of the above discussion, I have assumed that $\BK$
and $\BBd$ are both positive, as per theoretical expectations. If we
relax this assumption --- and I remind the reader that the original
excitement over measurements of CP violation in the $B$ system was the
ability to test the SM {\it without} theoretical input --- then this
has two consequences. First, the unitarity triangle can point up or
down. And second, the signs of the extracted functions $\sin 2\twida$,
$\sin 2\twidb$, $\cos 2\twidg$, etc.\ are uncertain.

Unfortunately, in this situation, one can no longer definitively
establish the presence of new physics. Consider again the case where
$(\alpha,\beta,\gamma) = (70^\circ,20^\circ,90^\circ)$, and $\thetaNP
= -50^\circ$. If $\sin 2\twida$, $\sin 2\twidb$ and $\cos 2\twidg$
have the ``wrong'' sign (i.e.\ $\BBd < 0$), then the two discretely
ambiguous solutions are
\begin{eqnarray}
(\twida_1,\twidb_1,\twidg_1) & = & (-20^\circ,-70^\circ,-90^\circ) ~, \nn\\
(\twida_2,\twidb_2,\twidg_2) & = & (-70^\circ,-20^\circ,-90^\circ) ~.
\end{eqnarray}
That is, we find the same solutions as before, except that they are
negative. Now, however, since (e.g.) $\cos 2\twidb$ also has the
``wrong'' sign, DAR will choose the $(\twida_2,\twidb_2,\twidg_2)$
solution. And if $\BK$ is allowed to be negative, then the unitarity
triangle points down, and this solution is completely consistent. In
other words, if we allow $\BK$ and $\BBd$ to be either positive or
negative, then we lose the ability to unambiguously conclude that new
physics is present. It is therefore important to try to experimentally
verify the signs of $\BK$ and $\BBd$.

\section*{Acknowledgments}
This talk was based on work done in collaboration with Boris Kayser. I
thank the organizers of PASCOS '99 for a stimulating conference. This
work was financially supported by NSERC of Canada.

\end{document}